# The Allen Telescope Array: The First Widefield, Panchromatic, Snapshot Radio Camera for Radio Astronomy and SETI

Jack Welch, Don Backer, Leo Blitz, Douglas Bock, Geoffrey C. Bower, Calvin Cheng, Steve Croft, Matt Dexter, Greg Engargiola, Ed Fields, James Forster, Colby Gutierrez-Kraybill, Carl Heiles, Tamara Helfer, Susanne Jorgensen, Garrett Keating, John Lugten, Dave MacMahon, Oren Milgrome, Douglas Thornton, Lynn Urry, Joeri van Leeuwen, Dan Werthimer, Peter H. Williams, Melvin Wright

Jill Tarter, Robert Ackermann, Shannon Atkinson, Peter Backus, William Barott, Tucker Bradford, Michael Davis, Dave DeBoer, John Dreher, Gerry Harp, Jane Jordan, Tom Kilsdonk, Tom Pierson, Karen Randall, John Ross, Seth Shostak

Matt Fleming, Chris Cork, Artyom Vitouchkine

Niklas Wadefalk, Sander Weinreb

*Abstract*—The first 42 elements of the Allen Telescope Array (ATA-42) are beginning to deliver data at the Hat Creek Radio Observatory in Northern California. Scientists and engineers are actively exploiting all of the flexibility designed into this innovative instrument for simultaneously conducting surveys of the astrophysical sky and conducting searches for distant technological civilizations. This paper summarizes the design elements of the ATA, the cost savings made possible by the use of COTS components, and the cost/performance trades that eventually enabled this first snapshot radio camera. The fundamental scientific program of this new telescope is varied and exciting; some of the first astronomical results will be discussed.

Manuscript received February 1, 2009. , This work was supported in part by the Paul G. Allen Family Foundation grant 5784, National Science Foundation grants 0540599, and 0540690, , generous contributions from Nathan Myhrvold, Greg Papadopoulos, Xilinx Inc., the SETI Institute, UC Berkeley, and other private and corporate donors. .

The first grouping of authors is with the Radio Astronomy Laboratory, University of California Berkeley (Jack Welch, phone: 1-510-643-6543, fax: 1-510-642-3411, e-mail: wwelch@astro.berkeley.edu). Joeri van Leeuwen is now with ASTRON in the Netherlands.

The second grouping of authors is with the SETI Institute, Mountain View, CA (e-mail: tarter@seti.org). William Barott is now with Embry-Riddle Aeronautical University in Daytona Beach, FL. Dave DeBoer is now the project manager for ASKAP with the ATNF CSIRO in Australia.

The third grouping of authors is with Minex Engineering in Antioch, CA (email: mcfmec@pacbell.net)

The final grouping of authors is with the California Institute of Technology in Pasadena, CA (email: sweinreb@earthlink.net). Nikolas Wadefalk is now with Chalmers University in Sweden.

## I. INTRODUCTION

THE Allen Telescope Array (ATA) is a "Large Number of Small Dishes" (LNSD) array designed to be highly effective for commensal surveys of conventional radio astronomy projects and SETI (search for extraterrestrial intelligence) targets at centimeter wavelengths. The importance of surveys in astronomy is well illustrated by the great successes of programs such as the Sloan Digital Sky Survey [1], and the ATA is planned to follow that example. It is well known [2] that for surveys requiring multiple pointings of the array antennas to cover a large solid angle in a fixed amount of time, the resulting point source sensitivity is proportional to ND, where N is the number of dishes, and D is the dish diameter, rather than $ND^2$, the total collecting area. Reasonable expectations for antenna and electronics costs then lead to the LNSD array as the optimum. The ATA will consist of 350 6m-diameter dishes when completed, which will provide an outstanding survey speed and sensitivity. In addition, the many antennas and baseline pairs provide a rich sampling of the interferometer *uv* plane, so that a single pointing snapshot of the array of 350 antennas yields an image in a single field with about 15,000 independent pixels. This number, the ratio of antenna beam width to array pattern beam width, is much smaller than the number of baselines and shows the large redundancy of the array. The goal is good image quality and high brightness sensitivity. Other important features of the ATA include continuous frequency coverage over 0.5 GHz to 10 GHz and four simultaneously available 600-MHz bands at the back-end which can be tuned to different frequencies in the overall band. Within these bands there are both 100-MHz spectral-imaging correlators and beamformers. The correlators have 1024 channels with adjustable overall bandwidths which permit high spectral resolution. Up to 32 separate beams may be formed to feed either SETI signal detectors or, for example, radio transient processors.

The ATA is a joint project of the SETI Institute in Mountain View, CA, and the Radio Astronomy Laboratory of the University of California, Berkeley. The initial design grew out of planning meetings at the SETI Institute summarized in the volume "SETI 2020" [3]. The design goals were (a) continuous frequency coverage over as wide a band as possible in the range 0.5 – 10 GHz for both SETI and conventional radio astronomy, (b) an array cost improvement approaching a factor of 10 over current array construction practices, (c) large sky coverage for surveys, (d) a collecting area as large as one hectare for a point source sensitivity competitive with other instruments, (e) interference mitigation capability for both satellite and ground based interference sources, and (f) both imaging correlator and beamformer capability with rapid data reduction facilities. An important realization at the planning meetings was that a very wideband inexpensive receiver could be built based on a MMIC chip that would have a very low noise temperature when cooled to only 60K (Weinreb, personal communication). A feed to accompany such a wideband receiver was clearly an important requirement, and some version of a log-periodic feed was an obvious choice. A simple cost optimization suggests that the antenna should cost about the same as the feed and receiver. An inexpensive antenna will be small, and the LNSD concept is a natural consequence. Finally, construction of large numbers of antennas and feeds that utilize commodity components and mass manufacturing techniques will lower the cost. A grant from the Paul G. Allen Family Foundation enabled the detailed hardware design and initial phase of array construction.

The ATA is now complete to 42 antennas and in operation. In the following sections, we describe the antenna design and operation, the feeds and receivers, the signal transport, the frequency adjustable bands, and the beamformers and correlators. We also discuss particular properties of the small array and the anticipated intermediate and final arrays. Highlights of the system are the frequency agility, the low background and sidelobes of the antennas, the wideband feed and input receiver, the analog fiber optical system, the large spatial dynamic range, the backend processing systems and the overall low cost.

The ATA is located at the Hat Creek Radio Observatory (HCRO) of the University of California Berkeley in northern California. Figure 1 shows an artist's conception of the anticipated 350-element array at the Observatory.

## II. THE ANTENNAS

The antenna is an offset Gregorian design that allows a larger secondary with no aperture blockage for good low frequency performance and also provides a clear aperture with lower sidelobes in the antenna pattern and lower thermal background. Having lower sidelobes is particularly important with the increasing level of satellite interference. The primary is an approximately 6m diameter section of a paraboloid, and the secondary is a 2.4m ellipsoid, four wavelengths across at the lowest operating frequency. The antenna schematic is shown in Figure 2(a). Also evident in Figure 2(a) is a metal shroud in darker tones, between the primary and secondary mirrors to protect the feed from interfering signals propagating along the ground and thermal noise from the ground. The shroud is approximately cylindrical. The radome cover is made from fabric as shown in the photo in Figure 2(b).

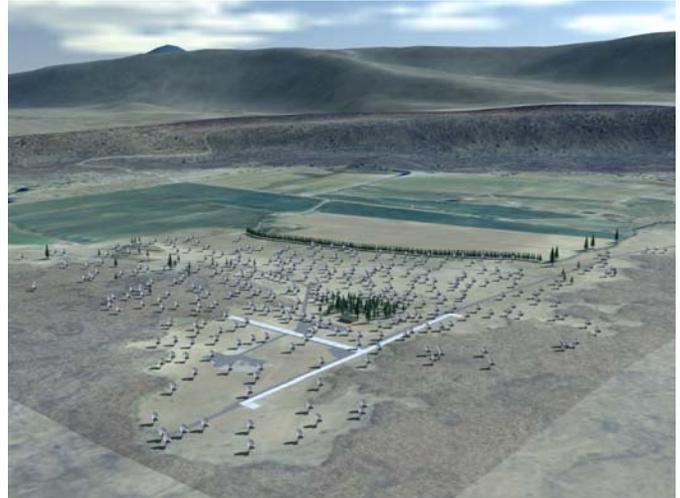

Fig. 1 Rendering of final ATA configuration with 350 antennas in the Hat Creek Valley of Northern CA.

The paper by Kildal [4] on diffraction losses and spillover for dual reflector systems served as a guide in the design. He gives the following expression for the aperture efficiency of such a system.

$$\eta_{ap} = \eta_{Ill} \left| 1 - C_b \left(\frac{d}{D}\right)^2 - C_d \sqrt{\frac{\lambda}{d}} \sqrt{1 - \left(\frac{d}{D}\right)^2} A_0 \right|^2 \quad (1)$$

Where $\eta_{Ill}$ is the feed illumination efficiency, $D$ is the primary diameter, $d$ is the secondary diameter, $A_o$ is the primary edge illumination by the feed, and $C_b$ and $C_d$ are geometric factors. (1) is the result of an analytic calculation that was verified by accurate simulations. The middle term is the primary blockage by the secondary. The third term gives the diffraction losses. Because we are using an offset Gregorian, we remove the geometrical blockage term. The edge diffraction losses remain as they are because they are essentially edge effects. For our choice of parameters, $C_d$ =0.454. The fractional edge illumination, $A_o$ =0.211 (that is, -13.5 dB), and the third term in (1) is just 0.048 $\lambda(m)^{1/2}$. We make one more plausible approximation that only one quarter of the diffraction losses couple to the ground because of the presence of the shroud. The formula then allows us to estimate the antenna gain and ground spillover. Knowing the detailed feed pattern, we can evaluate the aperture illumination and hence the antenna pattern. The edge diffraction gives the ground pickup contribution to the system temperature, 4K $f(GHz)^{-1/2}$. The most interesting result is the simple $\lambda^{1/2}$ dependence of the gain loss and ground spillover. This is a gradual dependence that is important in this application. Further details regarding the aperture and beam efficiencies and antenna patterns can be found in [5].

One more issue that remains is the possible interaction between the shroud and the feed. A stationary phase calculation [6] shows that the effect of the shroud as seen as a reflection at the input of the feed is given by the following formula.

$$|S_{11}| = \frac{\pi}{32} G^2 \frac{\lambda}{a} \qquad (2)$$

Where $a$ is the radius of the cylindrical shroud and $G$ is the gain of the feed at 90 degrees from its forward axis. The result is about -40 dB at the longest wavelength. echoes from the primary and secondary are also small.

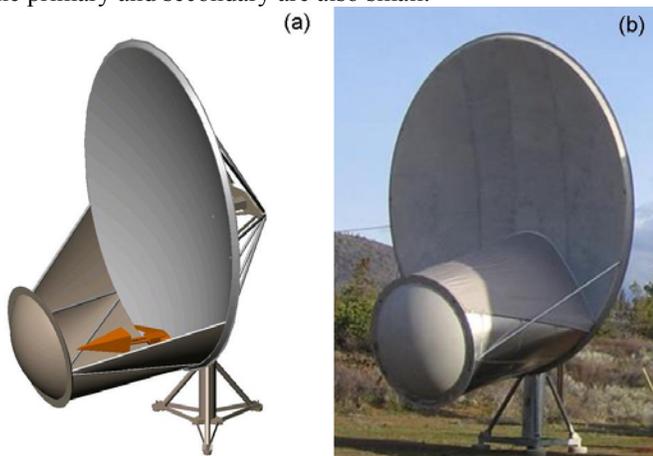

Fig. 2 The ATA offset Gregorian antenna, illustrating the shroud, feed placement, and radome cover.

The reflectors are made by Andersen Manufacturing (Idaho Falls, ID) employing the same hydroforming technology used to make low-cost satellite reflectors. The primary is a single formed piece of 3/16 inch thick aluminum with a thick stiffened rim, and is a patented, suspended design. The secondary is an 1/8 inch thick single aluminum piece supported at its rim. The pyramidal feed, also shown in Figure 2(a), is protected from the weather by a radio transparent radome cover shown in Figure 2(b) as well as by the shroud. Sidelobes of the mirror and feed system fall largely on the sky, and ground spillover is small. Antenna tipping measurements verify that geometrical optics spillover is small, except near zenith where it rises to 1% of the overall integrated gain function. The antenna has an az-el mount with a screw drive for the elevation and a special antibacklash azimuth drive. Slew speeds up to 3 °/s in azimuth and 1.5 °/s in elevation are possible.

The manufacturing process makes accurately reproduced mirrors. Photogrammetry was used in the development of the mold patterns for the hydroforming. Radio holographic measurements of the antenna structure show overall optical surfaces with total RMS errors of 0.7mm for night time observations. Figure 3 contains the radio patterns for two antennas made at 3.8 GHz; the similarities demonstrate the excellent reproducibility of the optics. The measured rms variation among the 11 antennas is 0.5%. In the middle of the day sunshine doubles the error to about 1.5 mm rms. Even though the surfaces have been bead-blasted to scatter the sunlight, the residual absorption of sunlight by the bare aluminum is enough to allow the differential heating that produces this degradation. The daytime surface rms is 1.5 mm,1/20 wavelength at our shortest operating wavelength, which is adequate.

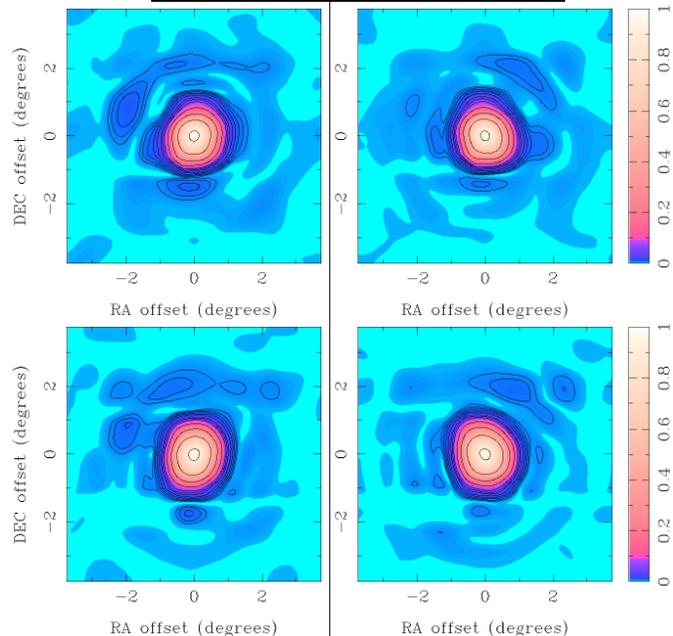

Fig. 3 Primary beam (power) pattern measurements of two representative ATA antennas in 2 polarizations at 3877 MHz; plotted on a logarithmic scale. Left images are x-polarization, right images are y-polarization. Contour lines are drawn at 0.9, 0.5, 0.25, 0.125, 0.0625, 0.04419, 0.03125, 0.015625, 0.00781, 0.00391, 0.00195, 0.000977 relative to the beam maximum. The average first sidelobe level is -30 dB; the small sidelobe peaks are -24 dB..

Further improvement in the daytime surface could be realized by painting it with white paint [7] and this upgrade may be carried out. Pointing accuracy is excellent during the night, typically around 10 arcsec, but degrades during the daytime. An insulated covering for the pedestal tower has been developed based on extensive testing and is scheduled to be installed shortly. Tracking and slewing are very smooth with trajectories based on a Kalman filtering scheme.

The antenna construction is designed for rapid and low cost manufacturing. The tower structure is a weldment. The alidade is a machined casting. A special welding machine makes all the tubular struts. It takes less than one half hour to stamp out a mirror. The final antenna assembly takes just eight person days.

## III. THE FEEDS AND RECEIVERS

The ATA feed is a pyramidal log-periodic feed (Figure 4-top panel) [8]. The motivation for the choice of this wideband feed was the development of very wideband low noise MMIC receivers [9]. The dual-linear-polarization feed pattern is optimum for illumination of an equivalent focal-length to antenna diameter of 0.65 which provides a fairly large depth of focus on the telescope. This patented feed incorporates a novel central metallic "pyramid" that allows low-noise amplifiers to be housed in a small cryogenic dewar placed directly behind the antenna terminals at the small feed end. Thus the cable losses are small which provides a low total effective receiver noise temperature. In transmission, the feed would excite a waveguide mode at the terminals that travels in the interior space between the arms and pyramid until it meets a resonant condition (at about λ/2 width of the feed). The wave is then radiated in free space back towards the small end. The linear dimensions of the feed yield an operating range from about 500 MHz to 10 GHz, and a linear drive allows the feed focus to be accurately set at the focal point of the two mirror system at any frequency. With the large focal ratio noted above, focusing for about 6.25 GHz allows the entire band to be accessed at the same time with only about 1-dB gain degradation at the band edges. The measured wide band width of the feed gain is shown in Figure 5. The feed gain was obtained from measurements of the transmission between the feed and several standard gain horns in an anechoic chamber. The gains of the horns are known to about 1% accuracy as are the cable losses, and corrections were made for the distances between the horn and the active zone of the feed. The gain variations in Figure 5 are real and are characteristic of log-periodic antennas [10]. The log-periodic nature of the feed is evident in the frequency dependence of its gain. Input reflections of the feed itself are measured to be < -14 dB over the operating band. Based on the measured feed gain and expected pattern the overall aperture efficiency is estimated to average 60% over the whole band.

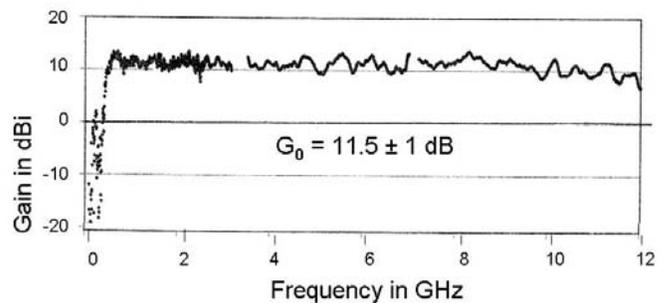

Fig. 5 Measured gain as a function of frequency for ATA feed.

The differential-input (active balun) low-noise amplifier is a monolithic microwave integrated circuit (MMIC) designed and packaged at Caltech [9]. The amplifier contributes approximately a (10+0.8f(GHz))K noise temperature across the band, roughly one third of the predicted total noise temperature when cooled to 60 K.

The expected value of the system temperature, $T_{sys}$, in Kelvins, including the individual noise source contributions is given by:

$$T_{sys} = \frac{4}{\sqrt{f}} + 6.3\sqrt{f} + 10 + 0.8f + 7 + 2.7 + \frac{3}{f^{2.7}} \quad (3)$$

$f$ is in GHz, the first term is due to diffractive spillover, the second term is the calculated Ohmic loss, the third and forth terms are attributable to the LNA, 7K is due to the atmosphere and geometric spillover, 2.7K is from the cosmic microwave background, and the last term represents synchrotron emission from the Galaxy. Figure 6 shows total $T_{sys}$ measurements for one of the receivers based on observations of the moon at the higher frequencies and Cass A for the lower frequencies. These were done with a single focus setting for 6.25 GHz, and the corrected values are shown. A constant 60% aperture efficiency was also assumed. The measured Tsys values near 2 GHz are lower than the predicted values, but the uncertainties are large at these frequencies. The agreement with the predictions, ~40K up to 5 GHz, is good, but the temperatures rise up to 80K above 8.5 GHz. It appears that the input losses are higher than predicted at the higher frequencies. The dashed red curve gives the results predicted by (3) and the solid red curve is the result of raising the assumed Ohmic loss by 50% to improve the fit. There is also fine structure in the system temperature that is the result of the real deviations of the feed gain from the 11.5 dB average used for the measurements. The results are quite good, in any case.

The major analog system components are the LNAs for the two polarizations, with gains ~30 dB, variable gain post-amplifier modules utilizing wide bandwidth RF Micro Devices gain blocks and Hittite variable step attenuators. The entire band, 0.5 - 10 GHz, is brought back to a centrally located processing facility via analog fiber-optic links developed for this project by Photonic Systems Inc (Burlington, MA) [12]. The post amplifiers and photonic modulators sit in a thermally controlled enclosure at the back of the feed. This space has its temperature regulated to 20 ± 0.1C. The photonic modulators and lasers have additional

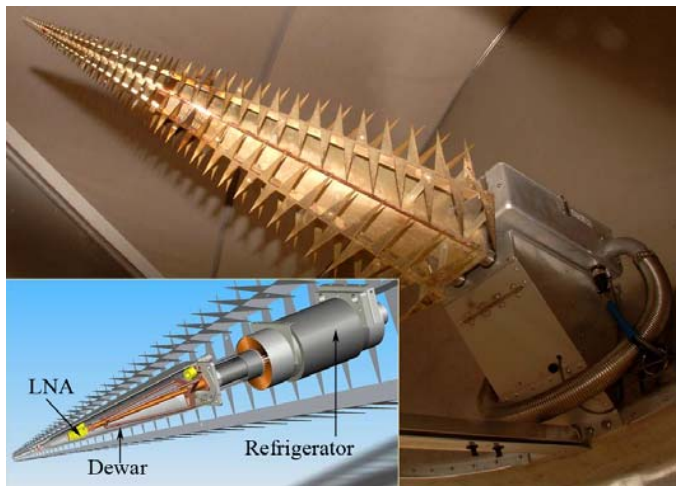

Fig. 4 log period feed mounted on linear focus actuator within the antenna; (inset) internal feed and receiver components

The feed geometry requires a pyramidal-shaped dewar with a glass window at its tip through which the balanced input lines from the feed terminals pass. The long narrow twin-leads of the dual polarizations connect the feed terminals to the balanced inputs of the low noise wide band balanced amplifiers (LNAs) in the vacuum dewar [9], as shown in the bottom panel of Figure 4. The LNAs in the dewar are cooled to about 60 K. The cryocooler, also shown in the bottom panel of Figure 4, is the Sterling Cycle cooler manufactured by STI [11]. Feed losses were calculated in the feed simulation [8]. Losses in the input transmission lines and the microstrip board at the tip of the feed were simulated.

temperature regulation of their own as required for the stability of the photonic system.

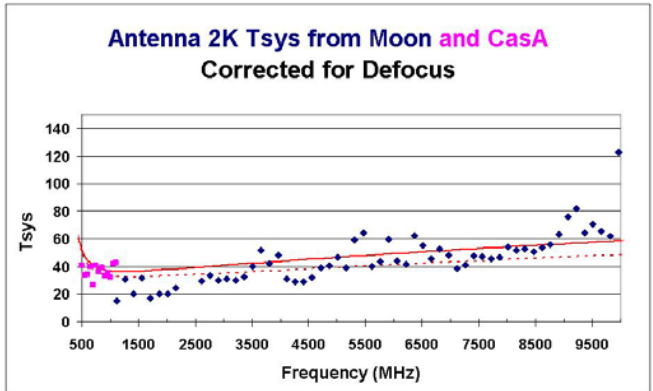

Fig. 6 Measured system temperature for one antenna using the moon and Casseopia A for reference, assuming 60% efficiency.

## IV. THE ARRAY

Figure 7 is an aerial photograph of the present array of 42 antennas. The construction tent is evident in the upper left of the picture, and the processor building is near the upper center of the picture. Figure 8a shows a snapshot distribution of baselines for the 42-antenna array. Figure 8b shows a snapshot of a 98-element array of baselines which is a possible next stage for the growth to the final 350. Figure 8c shows the baseline distributions for the final ATA-350. The azimuthal distribution of baseline angular directions for the ATA-350 is uniform, while the distribution of baseline lengths approaches a Gaussian as closely as possible for the ATA-350 (see Figure 8d). The final configuration of the ATA-350 is a nearly perfect Gaussian in the *uv* coordinates, except for the zero spacing. The Fourier transform of this distribution is the beam pattern, which is therefore also a Gaussian and an ideal point spread function.

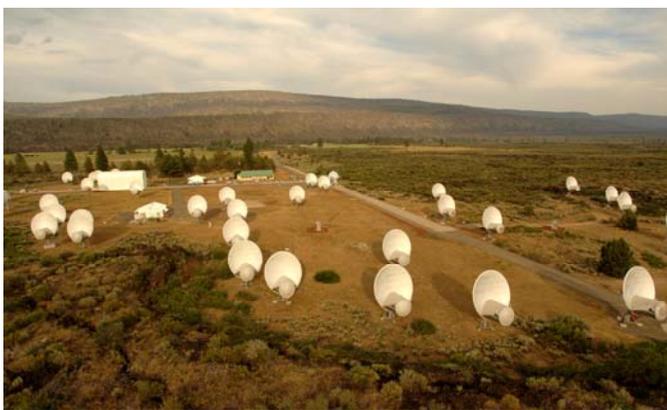

Fig. 7 Aerial view of the ATA-42 at the Hat Creek Radio Astronomy Observatory in northern CA.

### A. Antenna Nodes

In the center of Figure 7 one can see one of the node buildings, that typically serve 10 antennas. Underground PVC pipes connect the nodes to the processor building and to the antennas. The pipes are buried at a depth of about one meter. They contain the wide bandwidth optical fiber, multimode control fiber and the electrical mains. In addition, ambient temperature air is blown from each node to its antennas in order to provide adequate airflow over the electronics in the antennas. At a depth of one meter the ground temperature is approximately constant during the day at the mean daily temperature, and the ground pipe and ground serve as a heat exchanger to bring constant temperature air to each antenna.

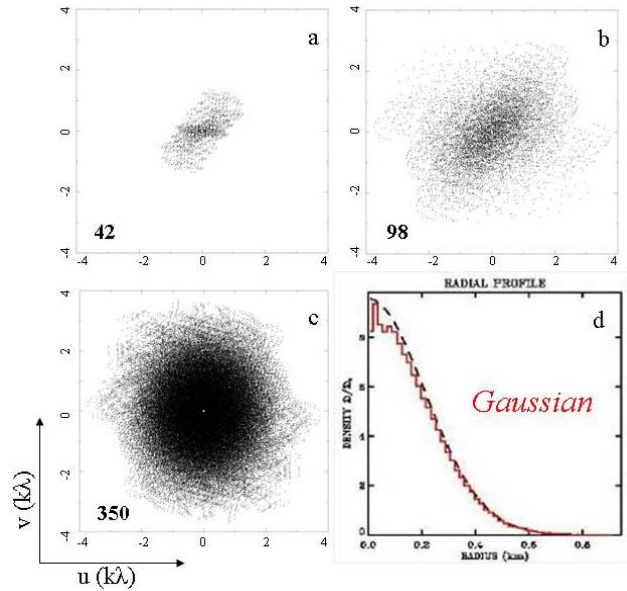

Fig 8 Snapshot of baseline distributions at 1.42 GHz and 10° declination for a) 42 antennas, b) 98 antennas, and c) 350 antennas. Panel d) is the distribution of baseline lengths for 350 antennas.

No further general heating or cooling of the air is needed. To understand the temperature variations with time and depth in the soil we can study the diffusion of heat into the ground from the surface. Measurements show that the surface temperature is approximately made up of four components, an annual average of about 287K and three periodic components: a daily component of 8C, an annual component of about 6.5C, and a weather component of about 5C with a typical period of about 12 days and a random phase. The diffusion into the ground is governed by the usual equation [13]

$$\frac{\partial^2 T}{\partial x^2} = \frac{C_v}{k}\frac{\partial T}{\partial t} \quad (4)$$

where $x$ is depth below the surface, $C_v$ is heat capacity, and $k$ is the thermal conductivity. The temperature distribution resulting from the four terms described above is:

$$T(x,t) = 287 + \sum_{n=1}^{3} T_{sp_n} e^{-x/\delta_n} \cos\left[\frac{2\pi t}{p_n} - \frac{x}{\delta_n}\right] \quad (5)$$

where the $T_{sp_n}$ are the three periodic surface temperatures, the diffusion depth is

$$\delta_n = \sqrt{\frac{k p_n}{\pi C_v}} \quad , \quad (6)$$

and the $p_n$ are periods in days. For n = 1, 2, 3, the periods are 1, 12, 365 days, and the corresponding e-folding diffusion depths in the average HCRO soil are 0.11, 0.38, and 2.1 m. The first two terms are the most rapid and therefore the most important and are the reason for the choice of one meter for the depth of the pipes. Temperature measurements verify the sub-surface distributions and the approximately constant temperature at this depth over the day. The mid-summer average temperature at this depth is about 22C. The mid-winter average temperature is about 7C.

Ambient temperature air at the node is blown into pipes leading to each antenna after passing through a filter. Reynold's number for the airflow in the pipes is about 20,000, insuring that there is effective turbulent heat transfer from the air into the pipe walls and thence to the ground. A heat flow calculation [14] shows that if $\Delta T_N(0)$ is the daily periodic air temperature component at the node, the temperature in the pipe at a distance $z$ down the pipe and at a later time $t$ is given by

$$\Delta T_N(z,t) = e^{-z/\gamma} \cos(2\pi t + \phi_0 z) \Delta T_N(0) \quad (7)$$

where

$$\gamma \propto F(m^3/s) \left| R_{turb} + R_{PVC} + Z_e \right| \quad , and \quad (8)$$

$$\varphi = \tan^{-1} \frac{\operatorname{Im}[Z_e]}{\left| R_{turb} + R_{PVC} + Z_e \right|} \quad (9)$$

$F$ is the air flow in cubic meters per second (about .06 m³/sec) as needed to cool components on the antenna. $R_{turb}$ is the thermal resistance of the heat flow to the pipe wall, $R_{PVC}$ is the thermal resistance of the PVC pipe, and $Z_e$ is the thermal impedance of the ground, a complex number because the ground has substantial heat capacity as well as conductivity. For average HCRO soil with 10% moisture, $\gamma \approx 16m$. Average pipe lengths are 30m, so that the typical residual temperature variation is about 1°C at the antenna. Note that any rapid environmental variations are completely filtered out.

## V. THE ANALOG ELECTRONIC SYSTEM

The remainder of the signal path lies within the processor building, which is near the center of the eventual 350-element array (see Figure 1). As noted above, the full 500 MHz to 10 GHz band of each polarization of each antenna is brought back to the processor building over single mode optical fiber as an analog signal. This signal is converted to baseband in an RF converter board (RFCB), see Figure 9. Figure 10 shows a system level block diagram of the analog and digital components of the ATA system. Within the RFCB, the fiber for each polarization goes to the optical diode in one of the two shielded input boxes. Four amplified parallel outputs from each box go to four first mixers. These mixers are fed by four different local oscillator signals labeled A, B, C, and D. Each mixer output passes through a passive bandpass filter that is 600 MHz wide, centered at 15.5 GHz. Tuning of the first LO over the range 16.3 to 26.6 GHz selects which 600-MHz wide input band of the 500 MHz to 10 GHz range passes through the filter. Following the filter is a second mixer, with a fixed LO set to 14.87 GHz, so that the filter bandpass is translated to 630 MHz ± 300 MHz. The selected band is further amplified and passed through a 200 MHz wide anti-aliasing filter, also centered at 630 MHz, before going onto the digital processing equipment. The bottom of the board shown in Figure 9 mirrors the top and provides signals from the other polarization. Because the four different first LOs are tuned separately, it is possible to perform simultaneous observations at four different frequencies. In order to avoid cross talk between signals originating at separate antennas, phase switching based on Walsh functions is applied to the second LO signals. This phase switching is removed following the digitization and before correlation or beamforming.

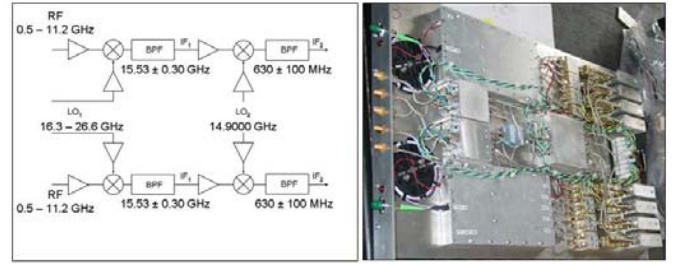

Fig. 9 Block diagram of one IF channel, dual polarization signal path through the RFCB, and RF converter board. Four dual-polarization IF channels are output.

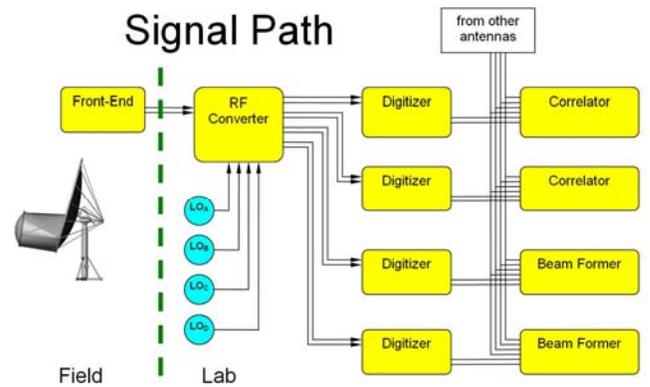

Fig. 10 – the current signal path for the ATA-42. A third beamformer will soon be added, with a goal of achieving 8 beamformers across the two LO tunings routinely used for SETI observations.

The current functioning rack of the 42 primary RFCB boards plus two spares, along with racks for two dual-polarization, spectral-imaging correlators and three beamformers are shown in Figure 11.

## VI. MONITOR AND CONTROL

The monitor and control system (M&C) integrates control of antennas, RF systems, correlators, and beam formers in a distributed Linux environment. It supports commensal telescope control by multiple "back-end" processes including astronomical image data collection, SETI spectrometry/detection, pulsar spectroscopy, etc. The system runs continuously with full remote-control support over the Internet. Approximately 30% of observations are initiated remotely.

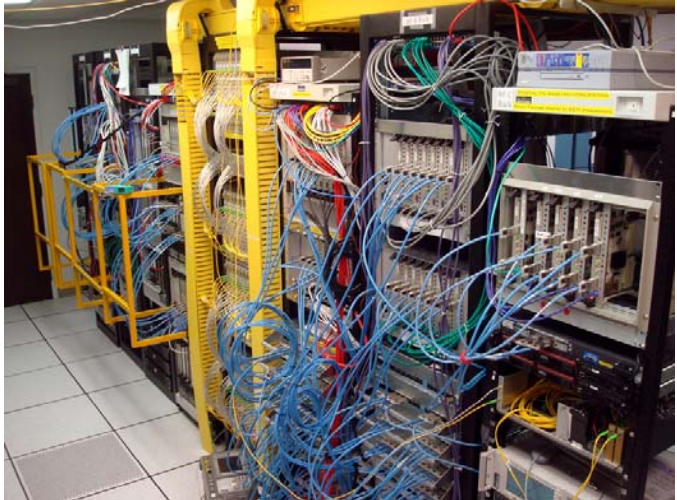

Fig. 11 left to right: 4 correlator racks, RFCB rack, 3 beamformer racks.

The M&C system is Java-based and runs on a handful of Linux servers that communicate via a proprietary message bus. Each antenna runs Java/Linux on a network booted single board computer (SBC) connected over fiber LAN to the control room on the same message bus. The SBC relays messages to six custom microcontroller boards to perform all the low level control of antenna position, focus, amplifier (LNA) settings, and monitoring functions.

## VII. Digital Signal Processors

The great flexibility of the ATA is enabled by multiple signal processing backends that can work simultaneously on the same, or different, independently tunable IF channels. At present the digital backends are spectral-imaging correlators, fast spectrometers for individual antennas, and beamformers feeding SETI signal processors and pulsar processors. More backends are expected in the future to accommodate different types of science. The analog to digital converters and subsequent field programmable gate array board for digital downconversion and fringe-rate/delay tracking are a development of the CASPER project [15] delivering eight bit quantization. There is precise level adjustment at the input of the ADC for each antenna in order to guarantee adequate dynamic range in the presence of radio Frequency Interference (RFI).

### A. Digital Correlators

The ATA correlator relies on an "FX" architecture that first divides the signal into frequency channels then cross-multiplies the result for each antenna pair. These products form the spectrum of the cross-correlation. A polyphase filter bank (PFB) [16] is used in order both to avoid cross-talk between channels and to lower the effects of narrow band RFI. With the eight-bit quantization and with the levels set so the background is just on scale, only signals greater than about 40 dB will overdrive the ADC. There is bit growth in the polyphase filter bank calculation, and the cross-correlation products must be limited to four bits. However, bit-selection can be made separately for each of the 1024 spectral channels, which will permit a large dynamic range in signal level The transposition of the filter bank channels is handled by a unique "corner turner" design [17].

The present correlators have an overall bandwidth of 100 MHz with 1024 spectral channels. Thus they make 1024 different spectral maps. The overall bandwidth can also be made smaller in steps of two for more spectral resolution. The 42-antenna array will produce snapshot maps with 250 independent pixels over the primary beam, but that number can be increased easily by a factor of a few using the traditional earth rotation synthesis.

### B. Beamformers

The other basic data processing scheme is the construction of beamformers. For this we add the outputs of the antennas phased up for a particular direction in the sky over the 100 MHz bandwidth. Figure 12 shows the cross-section of a uniformly weighted beam, and a dotted line that shows the beam with a null synthesized to avoid RFI.

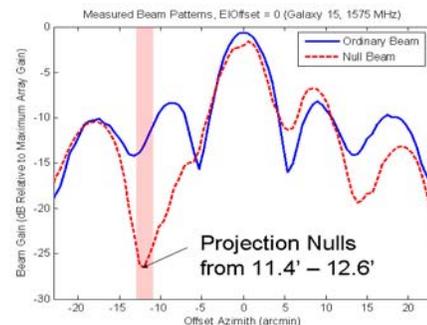

Fig. 12 Placement of a single -28 dB spatial null offset from the primary beam direction.

### B.1 The Prelude SETI System

The current generation of SETI signal detection hardware and software is called Prelude. It relies on rack mounted PCs that have been augmented by two custom accelerator cards based on DSP and FPGA chips. Each Programmable Detection Module (one of the 28 PCs) can analyze 2 MHz of dual-polarization input data to generate spectra with spectral resolution of 0.7 Hz and time samples of 1.4 seconds.

Figure 13 shows a "waterfall plot" of the amplitude in each 0.7 Hz channel as a function of time (increasing upwards); every observation produces 86016 such plots per polarization. The green lines have been inserted to draw the eye to a very faint track produced by the carrier signal from the Voyager 1 spacecraft, now just reaching the heliopause. Although it is barely visible to the eye, this signal is readily detected by efficient algorithms that add up the detected power along all possible drift paths in the frequency-time plot (see inset in Figure 13). During SETI observations, no attempt is made to compensate for the (mainly) diurnal acceleration of the Earth. A signal that was transmitted at a single tone, and pre-compensated to remove all its intrinsic acceleration relative to a geocentric reference frame, would therefore be detected with a Doppler drift rate of ~0.1 Hz/s by the Prelude system. Signals arriving with ≈ zero drift rate can therefore be discarded as interference (in fact the closer to zero the drift rate, the more likely the source of interference is to be locked to the observatory frequency standard, i.e. to be self-generated).

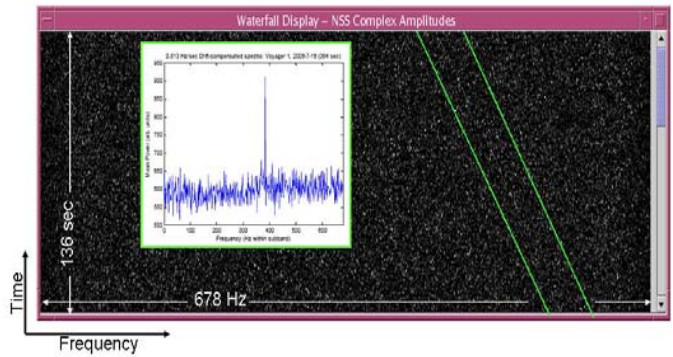

Fig. 13 Waterfall plot (frequency increasing to the right, and time increasing upwards) showing detection of carrier signal from Voyager 1 at a distance of 106 astronomical units. Green lines are intended as a guide for the eye. The inset shows the integrated signal, easily detected by the signal processing algorithms.

Current SETI signal detection algorithms search in near-real time for CW and narrowband pulsed signals, using two circular polarizations. Such signals will propagate through the ISM with little or no distortion, and to the best of our knowledge are not produced by any type of astrophysical emitter. A premium is placed on immediate recognition and classification of such signals, so that follow-up observing procedures can be automatically invoked to further investigate candidate ETI signals during the next observation. This has proven to be the most effective method of conducting efficient observations in a dynamically changing RFI environment.

The Prelude system will soon be replaced by SonATA (SETI on the ATA), a cluster of commodity servers with a full software-only implementation of the signal detection and automated follow-up processing now done by Prelude. This will make possible graceful future growth to accommodate more antennas, more bandwidth, and more signal detection algorithms for more complex classes of signals.

Because of the large field of view of the ATA, there are likely to be many of the ~250,000 currently identified SETI target stars within every pointing direction of the array. This enables a commensal observing strategy in concert with many of the large-scale astrophysical surveys planned for the ATA. Figure 14 shows a particular field that contains the galaxies M81 and M82, and the positions of the 50 stars that were recently observed pair-wise (to help with RFI excision) by Prelude operating at L-band.

*B.2 Pulsar Processor*

An initial implementation of a 100-MHz bandwidth dual–polarization pulsar processor has been built with FPGA-based components designed by the CASPER (Center for Astronomy Signal Processing and Engineering Research) group at Berkeley. Evolution of this design is now being deployed at a

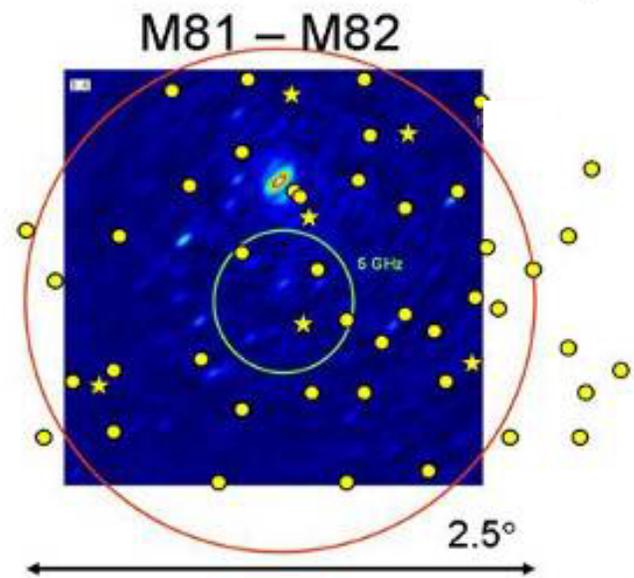

Fig14 – Target SETI stars observed while M81-82 complex was being mapped at L-band (1.42 GHz)

number of telescopes world wide. The input to the processor is the digital phased-array output of the beamformer. A 2048-point PFB algorithm, which is as described in section VII.A, is applied to resolve the dispersed pulsar signals. The resultant signals from orthogonal polarizations are cross-correlated, and a full set Stokes parameters are formed. Results are sent to an offline computer for storage and subsequent processing a kHz rates. Future signal processing will attend to voltage-based dedispersion for the high-precision time of arrival measurements.

*C. The Fly's Eye System*

In this third digital system currently deployed on the ATA-42, all the antennas are pointed in slightly different directions with de-dispersion spectrometers on each looking for sharp background pulses over a wide field. A recent use of this backend system to search for rare, highly dispersed, transient pulses is described by Siemion *et al* [18].

VIII. EARLY IMAGING SCIENCE RESULTS

To demonstrate the wide field capability of the array, we have made a large mosaic of continuum radiation at a frequency of 1.4 GHz. The mosaic is the combination of 1-minute integrations on 350 separate pointings producing an image that is 800 square degrees in size, or 2% of the sky. Several thousand sources are detected. These sources are primarily relativistic jets associated with massive black holes in galaxies at very large distances from the Sun. In Figure 15, we show a subset of this image made from 10 pointings, approximately 25 square degrees in size. A single observation of the 800 square degree field can be made in 8 hours. Repeated observations of the large field are made to survey for transient and variable sources.

Two of the bright spiral galaxies in the Local Group, M31 (Andromeda) and M33, are imaged in Figure 16 in the emission line of neutral hydrogen. The resolution is shown by the small ellipses in the lower left hand corners. Background continuum sources have been removed from the images. The image of the Andromeda galaxy (on the left) shows the familiar ring of atomic gas that lies beyond the optical image. These objects are quite extended on the sky and difficult to image with conventional arrays of large antennas.

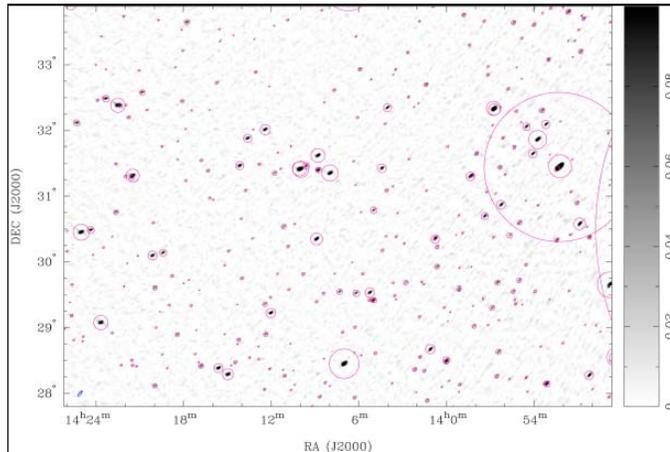

Fig. 15  1.4 GHz continuum mosaic of 25 square degrees of the sky (S. Croft et al. 2009, in prep.). The grayscale wedge shows 1.4 GHz flux density in Jy as measured by the ATA, and the circles (scaled in proportion to flux density) show the positions of sources from the NRAO VLA Sky Survey (J. Condon et al. 1998, AJ, 115, 1693). The match in position and flux density between the two surveys is good.

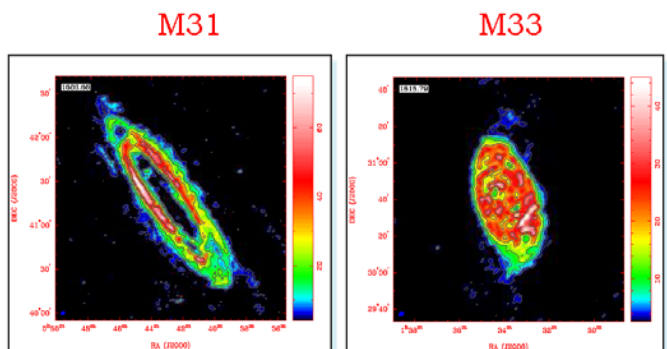

Fig. 16  Local group galaxies M31 and M33 mapped in the HI line

## IX. Further developments

In addition to the planned build out of the array to 350 antennas and construction of more correlator and beamformer backends, improvements in the design of the feed are being considered. The plan is to carry the input end to a smaller cross-section and cool the last 4 inches of the feed tip to 60K. The goal is an overall system temperature below 40K up to 24 GHz.

## Acknowledgment

During the initial technology development phase of the ATA project, the ATA Technical Advisory Panel served as an excellent source of guidance and information, and we would like to acknowledge the contributions of Barry Clark (NRAO), Howard Davidson (Sun Microsystems), Ron Ekers (ATNF), James Gosling (Sun Microsystems), Paul Horowitz (Harvard), Peter Napier (NRAO), John O'Sullivan (ATNF), Pierre St.-Hilaire, Lou Scheffer (Cadence, Inc.), Arnold van Ardenne (ASTRON). Additional assistance was provided by Ken Johnston (USNO).

## References

[1] J. Gunn et al, "The 2.5m telescope of the Sloan Digital Survey", Astrophysical Journal, vol 131, p2332, 2006.  Adelman-McCarthy, J.K. et al, "The Sixth Data Release of the Sloan Digital Sky Survey", Astrophysical Journal Supplements, vol 175,  p 297, 2008.
[2] Sargent, A. I., and Welch, W. J. "Millimeter and submillimeter interferometry of astronomical sources", ARAA vol 31, p297s, 1993.
[3] *SETI 2020. A Roadmap for the Search for Extraterrestrial Intelligence* Managing Editor, Seth Shostak; Editors: Ronald D. Ekers, D. Kent Cullers, John Billingham, and Louis K. Scheffer,2002, SETI Press, Mountain view, CA.
[4] P.-S. Kildal, "The effects of subreflector diffraction on the aperture efficiency of a conventional Cassegrain antenna-An analytical approach", IEEE Trans. Antennas Propagation, vol.AP-31, No. 6, pp 903-909, Nov.1983
[5] W. J. Welch and D. R. DeBoer, ATA memo # 66. " Expected properties of the ATA antennas" July 18, 2004, http://ral.berkeley.edu/ata/memos/memo66.pdf.
[6] W. J. Welch, ATA memo # 28, "Interactions between the LP feed and the shroud and reflectors", 2001, http://ral.berkeley.edu/ata/memos/memo28r1.pdf.
[7] William Imbriale, private communication, 2008.
[8] G. Engargiola and Wm. J. Welch, "Log-periodic antenna"  US Patent #6,677,913 (2004)
[9] N. Wadefalk and S. Weinreb, ATA memo #67 "Noise testing for the active and passive baluns for the ATA., 2004, http://ral.berkeley.edu/ata/memos/memo67.pdf
[10] V.H. Rumsey, *"Frequency Independent Antennas"*, Academic Press, 1966.
[11] Superconductor Technologies Incorporated, Santa Barbara, CA 93111.
[12] E. Ackerman, C. Cox, J. Dreher, M. Davis, and D. DeBoer, "Fiber-Optic Antenna Remoting for Radio Astronomy Applications" URSI 27th General Assembly,  Maastricht (2004)
[13] H. S. Carslaw and J. C. Jaeger, *Conduction of Heat in Solids*, Oxford University Press, Oxford UK, 1946
[14] W.J. Welch, ATA memo #68,"Some Thermal Issues Regarding the ATA Node Air System", 2005, http://ral.berkeley.edu/ata/memos/memo68.pdf
[15] A. Parsons, D. Backer, C. Chang, D. Chapman, H. Chen, P. Crescini, C. de Jesus, C. Dick, P. Droz, D. MacMahon, K. Meder, J. Mock, V. Warzynek, D. Wertheimer, M. Wright, "PetaOp/Second FPGA Signal Processing for SETI and Radio Astronomy." *Proceedings of  the Asilomar Conference on Signals and Computers,*  November 2006, Pacific Grove. CA.  http://casper.berkeley.edu/
[16] W. L. Urry, M. Wright, M. Dexter, D. MacMahon, ATA memo # 73, "The ATA Correlator", 2000, http://ral.berkeley.edu/ata/memos/memo73.pdf.
[17] W. L. Urry, ATA memo #14, "A Corner Turner Architecture", 2000, http://ral.berkeley.edu/ata/memos/memo14.pdf..
[18] Siemion, A., et al., "New SETI Sky Surveys for Radio Pulses," astro-ph/0811.3046, submitted November 2008, Acta-Astronautica, http://arxiv.org/abs/0811.3046